\documentclass[11pt,twoside]{article}

\usepackage{prcsa2008}
\usepackage{epsf}
\usepackage{psfig}
\usepackage{lscape}
\usepackage{graphicx}

\begin{document}
\title{THE FORMATION, EVOLUTION AND PARAMETERS OF SHORT-PERIOD LOW-MASS X-RAY
BINARIES WITH BLACK-HOLE COMPONENTS}   
\author{L.R. Yungelson}   
\affil{Institute of Astronomy, Russian Academy of Sciences, Moscow, Russia}    
\author{J.-P. Lasota}   
\affil{IAP, UMR 7095 CNRS, UPMC Univ Paris 06, Paris,
           France\\
           Astronomical Observatory, Jagiellonian University, Krak\'ow, Poland}    
\begin{abstract} 
We discuss the formation, evolution and observational
parameters of the population of short-period
($\raise-.5ex\hbox{$\buildrel<\over\sim$}10$\,hr)
 low-mass black-hole binaries (LMBHB). Their evolution is
 determined by the orbital angular momentum loss and/or
nuclear evolution of the donors.  All observed semidetached
LMBHB are observed as soft X-ray transients (SXTs). The absence
of observed short-period stable luminous X-ray sources with
black holes and low-mass optical components suggests that upon
RLOF by the donor, the angular-momentum losses are
substantially reduced. The model with reduced angular-momentum
loss reasonably well reproduces the masses and effective
temperatures of the observed secondaries of SXTs. Theoretical
mass-transfer rates in SXTs are consistent with those deduced
from observations only if the accretion discs in LMBHB are
truncated. The population of short-period LMBHB is formed
mainly by systems which at RLOF had unevolved or slightly
evolved donors (abundance of hydrogen in the center $X_c\,
\raise-.5ex\hbox{$\buildrel>\over\sim\,$} 0.35$). Our models
suggest that a very high efficiency of common envelopes
ejection is necessary to form LMBHB.
\end{abstract}


\section{Introduction}
Currently, ten Galactic dynamically-confirmed black-hole
candidate X-ray binaries with K/M spectral type secondaries and
orbital periods $\raise-.5ex\hbox{$\buildrel<\over\sim$}$1 day
have been observed \citep{2006csxs.book..157M}. All these
objects are SXTs. Their X-ray luminosity may vary by 5--8
orders of magnitude between quiescence and the peak of the
outburst. Recurrence times spread from a about a year to tens
or years and could be even longer. Their variability is
interpreted as resulting from a thermal-viscous instability of
irradiated accretion discs around black holes in semidetached
binaries \citep[see][and references
therein]{2001NewAR..45..449L}. The estimated number of low-mass
black-hole binaries (LMBHB) in the Galaxy ranges from several
hundred to several thousand
\citep{1997ApJ...491..312C,1998A&A...333..583R}. The list of
short-period black-hole SXTs and some of their observed and
inferred parameters are rendered in Table 1. Observational data
presented in Table 1 is based on the survey of the literature
\citep[see ][]{2008arXiv0802.4375Y}.

Below, we consider the formation, evolution and some
observational properties of  LMBHBs.

\begin{table}[!ht]
\caption{Known short-period black-hole SXTs. }
\smallskip
\begin{center}
{\small
\begin{tabular}{lllccl}
\tableline
\noalign{\smallskip}
No. & Name  & $P_{\rm orb}$,  & Sp & $\langle \dot{M}_{\rm recc} \rangle$ & $~~\dot{M}_{\rm in}$, \\
&&hr&&${\rm M_{\odot}~yr^{-1}}$&${\rm M_{\odot}~yr^{-1}}$ \\
\hline
1. & XTE J1118+480 (KV UMa)   & 4.10  &  K5-M1  & $1.9 \times 10^{-12}$   & $3.0 \times 10^{-10}$  \\
2. & GRO J0422+32 (V518 Per)  & 5.09  & M2-M4   & $1.3 \times 10^{-11}$   & $1.9 \times 10^{-10}$  \\
3. & GRS 1009-45 (MM Vel)     & 6.84  & K7-M0.5 & $4.4 \times 10^{-11}$   & $1.8 \times 10^{-10}$  \\
4. & XTE J1650-500            & 7.68  & K4      &  $2.0 \times 10^{-11}$  & $2.8 \times 10^{-10}$  \\
5. & A0620-00 (V616 Mon)      & 7.75  & K3-K7   & $3.3 \times 10^{-11}$   & $2.8 \times 10^{-10}$  \\
6. & GS 2000+25 (QZ Vul)      & 8.28  & K3-K6   &  $2.0 \times 10^{-10}$  & $5.5 \times 10^{-10}$  \\
7. & XTE J1859+226 (V406 Vul) & 9.12  & G5-K0   & $4.1 \times 10^{-10}$   & $1.0 \times 10^{-9}$  \\
8. & GRS 1124-68 (GU Mus)     & 10.39 & K3-K7   & $3.4 \times 10^{-10}$   & $3.5 \times 10^{-10}$  \\
9. & H 1705-25 (V2107 Oph)    & 12.50 & K3-K7   & $5.5 \times 10^{-11}$   & $4.1 \times 10^{-10}$  \\
10. &  4U 1543-47  (IL Lup)   & 27.0  &  A2     & $4.2 \times 10^{-10}$   & $1.3 \times 10^{-9}$  \\
\noalign{\smallskip}
\tableline
\end{tabular}
}
\end{center}
\vspace{-2mm} {\small Note:  $\langle \dot{M}_{\rm recc}
\rangle$ -- mass-transfer rate estimate based on recurrence
times, $\dot{M}_{\rm in}$ -- upper limit to mass-transfer rate
based on assumption of maximal truncation of accretion discs in
SXTs. }

\label{tab:pte}
\end{table}

\section{The model}
To obtain a model of the population of LMBHBs one needs  to
follow two steps: (i) the time-dependent
formation of the population
of detached ``black-hole + main-sequence star" binaries
 and
(ii) the subsequent evolution of every binary till the Hubble
time.

The threshold for the masses of black-hole producing binary
components
is (20 -- 25)$\rm M_{\sun}$ \citep[e.g.,
][]{1998A&A...331L..29E}. The masses of the secondaries in SXTs
suggested by their spectral types are
$\raise-.5ex\hbox{$\buildrel<\over\sim$}1.5\rm M_\odot$.
Therefore the sequence of transformations of a binary which
results in the formation of a LMBHB may be the following:
\begin{trivlist}
\item[~(a)] the primary component evolves off main-sequence and
    becomes a supergiant,
\item[~(b)] the supergiant overflows Roche lobe and forms a
    common
    envelope, since mass ratio of components  $q\gg1$; if the
    components do not merge, the primary becomes a
    Wolf-Rayet (WR) star,
\item[~(c)] the WR star explodes as a supernova and forms a
    black
    hole.
\end{trivlist}

If the binary is not  disrupted by the supernova explosion, a
system with a black hole accompanied by a low-mass
main-sequence star, i. e., a LMBHB,  is formed. If the
separation of components is sufficiently small
($\raise-.5ex\hbox{$\buildrel<\over\sim$}10\rm R_\odot$), the
orbital angular momentum loss (AML) via magnetic braking (MB)
and/or gravitational wave radiation (GWR) may bring secondary
component of the system to the Roche-lobe overflow (RLOF).

The range of post-common-envelope binary separations which
allow the formation of  semidetached LMBHBs is very narrow.
Hence the probability of formation of a SXT depends very
strongly on the physical processes that determine the
semi-major axis of the binary, i.e. the stellar winds at all
stages of evolution and, especially, the  efficiency of
ejection of the common envelope. For the latter, the ratio of
final $a_f$ to initial $a_i$ separations of components is equal
to \citep{1984ApJ...277..355W}
\begin{equation}
\label{eq:ce}
\frac{a_f}{a_i}= \frac{M_{1,c}}{M_1}\left[ 1+\left( \frac{2}{\alpha_{ce}\lambda r_{1,L}}\right)
\left( \frac{M_1-M_{1,c}}{M_2}\right) \right]^{-1},
\end{equation}
where $\alpha_{ce}$  is  the common envelope ejection
efficiency, $\lambda$\ -- the parameter of the binding energy
of the stellar envelope, $M_1  $ and $ M_{1,c} $ are the
initial mass of the mass-losing star and the mass of its
remnant, $r_{1,L}$ is the dimensionless radius of the star at
the beginning of mass transfer, $ M_2 $  is the mass of
companion. If $M_1 \gg M_2$,  then,  crudely, $a_f/a_i \propto
\alpha_{ce}\lambda$. Both terms in the latter expression are
highly uncertain \citep[see discussion
in][]{2003MNRAS.341..385P,2006MNRAS.369.1152K,2008arXiv0802.4375Y}.
Indirect estimates involving formation scenarios for binaries
with neutron-star or black-hole components suggest
$\alpha_{ce}\lambda \raise-.5ex\hbox{$\buildrel<\over\sim$} 2$.
This might mean that sources other than the orbital energy are
involved in the ejection of the common envelope, though, the
nature of these sources is still not fully understood. In
practice, $\alpha_{ce}\lambda$ remains a parameter, tuning of
which allows reproducing the properties of specific systems or
stellar populations.

In the second step of modeling, the evolution of every system
is followed taking into account the AML via gravitational wave
radiation and/or magnetic stellar wind (MSW) and, if necessary,
nuclear evolution of the main-sequence star.

\section{The galactic population of LMBHBs}

In \citet{2006A&A...454..559Y} and \citet{2008arXiv0802.4375Y}
was carried out the population synthesis for galactic LMBHBs
for three values of $\alpha_{ce}\lambda $: 0.1, 0.5 and 2. We
refer the reader to these papers for the details of the
computations. Two main conclusions were drawn from the models
computed  in these studies.

First,
the model for $\alpha_{ce}\lambda=$0.1 is not compatible with
observations of SXTs since in this case the overwhelming
majority of black holes has masses
$\raise-.5ex\hbox{$\buildrel>\over\sim$} 14\,M_{\odot}$,
exceeding the largest estimated black hole mass in known SXTs
\citep[$9.7\pm0.6$ for A0620-500, see][]{2007ApJ...663.1215F}.

Second,
if the AML via MB
 for LMBHBs is treated in
a ``standard'' way, assuming after \citet{1981A&A...100L...7V}
that for components of close binaries the braking law for
single field stars \citep{1972ApJ...171..565S} can be
extrapolated over the range of rotational velocities from
several 10 to several 100 km/s and that the spin-orbit coupling
is efficient,  model mass-transfer rates for LMBHBs at $P_{ \rm
orb} \raise-.5ex\hbox{$ \buildrel>\over\sim$} 2$\,hr are so
high that these systems might have stable hot discs according
to the disc instability model (DIM) criterion of
\citet{1999MNRAS.303..139D}. But such stable and bright LMBHBs
have not been observed.

A strong reduction of the magnetic braking efficiency in close
binaries as compared to single stars is suggested also by the
data on stellar rotation in young open clusters
\citep{2002ASPC..261...11C,2003ApJ...582..358A} and the mass
transfer rates in cataclysmic variables \citep[e.g.,
][]{1988MNRAS.231..535H,2003ApJ...599..516I}. In
\citet{2006A&A...454..559Y} and \citet{2008arXiv0802.4375Y} we
computed a  population model of LMBHBs assuming that MB stops
operating once the RLOF occurs (``no-MB'' model). We found that
the number of galactic LMBHBs remains of the same order of
magnitude as in the case with active MB (several 1000) but all
systems are transient.

In \citet{2006A&A...454..559Y} and \citet{2008arXiv0802.4375Y}
we compared the $\alpha_{ce}\lambda=$2 ``no-MB'' model with
observations. Below, we present the model for
 $\alpha_{ce}\lambda=$0.5. In this case, the number of LMBHBs that reach contact
in Hubble time and evolve to $P_{ \rm orb} \raise-.5ex\hbox{$
\buildrel<\over\sim$} 10 $\,hr is $\approx 6000$ and $\approx
3000$ of them have currently $P_{\rm orb} \raise-.5ex\hbox{$
\buildrel>\over\sim$} 1.5 $\,hr.\footnote{Systems with shorter
$P_{\rm orb}$ have $q\raise-.5ex\hbox{$
\buildrel<\over\sim$}0.02$ and the character of mass exchange
in them is unclear because of the effect of resonances, see
\citet{2006A&A...454..559Y} for details.}  For comparison, in
the model for  $\alpha_{ce}\lambda = $ 2 these numbers are
about 12000 and 5000, respectively.

\begin{figure}[!t]
\plottwo{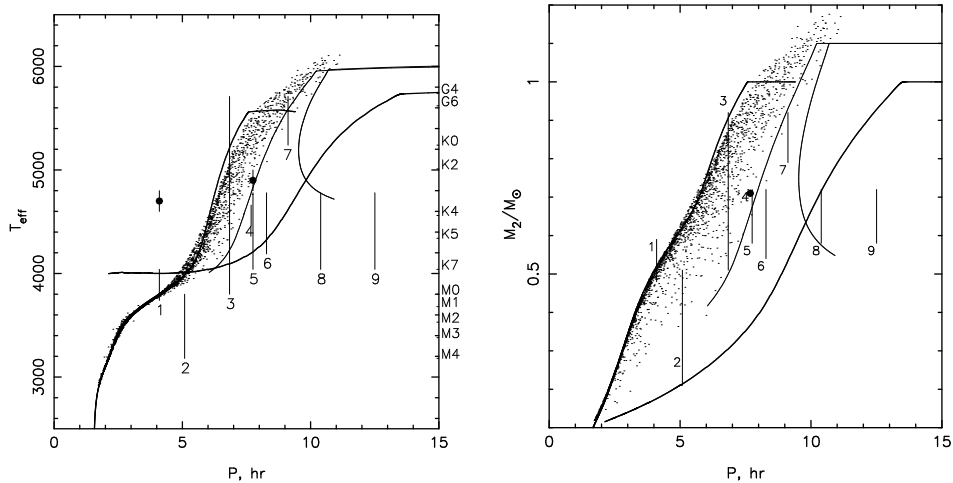}{yungelson_fig1b.eps}
  \caption[]{\textit{Left panel}: Model population (dots) vs. observational estimates of the
   ranges of effective
   temperatures of donors in SXTs inferred from their spectral types (vertical lines).
   The latter are annotated according to their number in
   Table~\ref{tab:pte}. Large filled circles give $T_{\rm eff}$ of donors derived
   from the fits to synthetic spectra. Heavy solid lines to the left and right,
   respectively, show
the approximate limits of the region which may be occupied by LMBHBs, while
thin solid lines show tracks for the donors in $(1.1+4)\rm M_\odot$ systems
which have at RLOF $X_c=$0.45 and 0.37,
respectively (see the text for discussion).
   The $Sp-T_{\rm eff}$ relation used in the paper is shown at the right
   border of the coordinate box.
\textit{Right panel}: Masses of donor-stars in modeled population. Vertical
lines show the ranges of $M_2$ inferred from their spectral types (annotation
like in the left panel). Heavy and thin solid lines -- the same tracks as
in the left panel.}
    \label{fig:pte}
\end{figure}

\subsection{Effective temperatures and masses of secondaries in LMBHBs}

When spectral types of the secondaries in observed SXTs are
known, one can evaluate their effective temperatures and masses
and to compare them with model predictions. Inevitably, such a
comparison may be only crude, since the spectra of the
secondaries are contaminated by the radiation of accretion
disks and hot spots and we have at our disposal $Sp-T_{\rm
eff}$ and $Sp-T_{\rm eff}$  relations for main-sequence stars
only \citep[relations from][were used]{2000asqu.book.....C}. In
the left panel of Fig.~\ref{fig:pte} we plot the effective
temperatures of donor-stars in modeled systems as a function of
orbital period and compare them with the ranges of $T_{\rm
eff}$ inferred from the spectral types of donors in SXTs.
 Having in mind uncertainties in the spectral type determinations and
$Sp - T_{\rm eff}$ scale, we find that the model satisfactorily reproduces
 $T_{\rm eff}$ of the donors in the LMBHBs with
$P_{\rm orb} \raise-.5ex\hbox{$ \buildrel<\over\sim$} 9$ hr.

In our models the efficiency of magnetic braking is 0. In
reality some MB might be acting as, probably, in the case of
cataclysmic variables \citep[see, for instance, ][but also
\ref{mtr} below]{2003ApJ...599..516I}. We show in Fig.
\ref{fig:pte} two ``limiting '' tracks: for the system with
initial masses of components 1 and 12 $\rm M_{\sun}$ and
post-circularization period $P_{\rm orb,0}=0.4$ day in which
the donor is almost unevolved at RLOF and MB does not act after
RLOF and for a (1+4)\,$\rm M_{\sun}$, $P_{\rm orb,0}=1.9$ day
system in which at RLOF donor has hydrogen abundance in the
center $X_c\simeq10^{-4}$ and the MB continues to operate
during mass-transfer with an efficiency corresponding to
\citet{1981A&A...100L...7V} law. Crudely, if the efficiency of
MB is not 0, the model population must be located between these
two limiting curves. Of course there will be a contribution
from lower and higher mass systems, as we plotted the 1\,$\rm
M_{\sun}$ tracks for simplicity only. Adding some MB to our
model will shift the population to the right, providing a
better agreement with observations.
Adding moderate MB will influence mainly the long-period
systems, without producing stable systems.

A similar satisfactory agreement of model populations with
observations is found for the masses of secondaries of SXTs
(Fig.~\ref{fig:pte}, right panel).

We note that, since for large initial $q$ the transformation of
separation of components in common envelopes depends linearly
on $\alpha_{ce}\lambda$, the distributions of model populations
in $P - T_{\textrm eff}$ and $P - M_2$ plots are similar for
$\alpha_{ce}\lambda$=0.5 and 2, and  only differ by the density
of points per unit area of the diagrams.

There are SXTs with $P_{\textrm orb}\approx$ (8--12) hr --
GRS~1009-45, XTE~1650-500, A0620-00, and GS~2000+25, which,
apparently, are located below the ``populated'' area in
Fig.~\ref{fig:pte}. However, we should note that in our
modeling we tried to avoid the effect of bifurcation of
evolutionary tracks --  evolution of systems to shorter or
longer periods upon RLOF, depending on the extent of hydrogen
depletion in the cores of the models. For this reason  we
restricted ourselves to systems evolving to shorter $P_{\rm
orb}$ only. However, our computations show that, if GWR is the
sole sink of angular momentum, binaries with $M_1 \approx
(4-12)\,\rm M_{\sun}$ and $M_2 \raise-.5ex\hbox{$
\buildrel<\over\sim$} 1\,\rm M_{\sun}$ evolve to longer periods
if at the instant of RLOF their secondaries have
 $X_c \raise-.5ex\hbox{$ \buildrel<\over\sim$} 0.4$. This is illustrated
in Fig.~\ref{fig:pte} by evolutionary tracks for initially
(1.1+4)\,$\rm M_\odot$ systems in which secondaries have,
respectively, $X_c \approx$ 0.45 and 0.37 at the RLOF. The
latter binary spends several Gyr evolving to longer periods.
Initial (post-circularization of the binary after supernova
explosion) orbital periods of these two binaries $P_0$ differ
by 0.1 day only. Since the distribution of the binaries over
$P_0$ is continuous, this proximity of initial parameters of
the binaries and striking difference in their evolutionary
behavior suggests a possibility of explaining the origin and
parameters of SXTs with $P_{\rm orb} \approx (8-12)$\,hr. This
conjecture has yet to be confirmed by detailed modeling.

\subsection{Mass-transfer rates in LMBHBs}
\label{mtr}

\begin{figure}[!t]
\begin{center}
\scalebox{0.5}{\rotatebox{-90}{\plotone{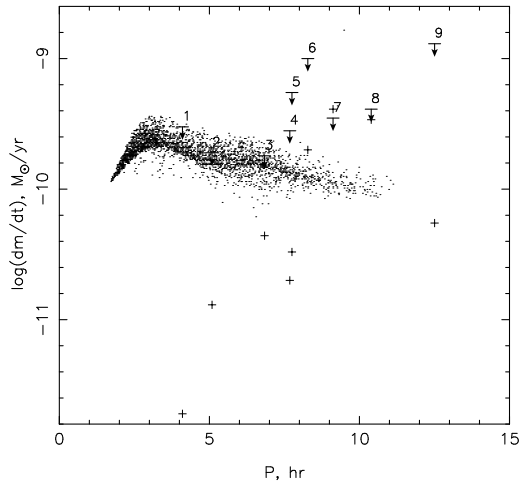}}}
   \caption[]{ Mass-transfer rates in model LMBHBs as a function of their orbital periods (dots).
Arrows mark
upper limits to the estimates of mass-transfer rates in observed
SXTs as given by Eq. (\ref{eq:dmdtmax}). Crosses are estimates of
mass-transfer rates in SXTs based on recurrence times.}
\label{fig:pdmdt}
\end{center}

\end{figure}

An estimate of mass-transfer rate in an SXTs may be obtained by
dividing the mass accreted during outburst by the recurrence
time.  However, (i) recurrence times are known for A0620-00
(about 60 yr) and 4U~1543-47 (about 10 yr) only;  (ii) it is
not evident that the rate calculated this way represents the
secular value, and  (iii) this method of estimate assumes that
between outbursts, when accretion disc is ``refilled",
accretion onto black hole does not occur. The last assumption
is put in doubt both by observations \citep[see e.g.][ and
references therein]{2007A&ARv..15....1D} and models
\citep[see][ and references therein]{2008arXiv0801.0490L} which
suggest that quiescent discs in SXTs are truncated and
therefore leaky. In the latter case, mass-transfer rate cannot
be larger than the critical-for-stability accretion rate at the
truncation radius \citep[see ][ for
details]{2006A&A...454..559Y}. The actual mass-transfer rate
should be between the values estimated by the two methods.

Using the expression for the critical accretion rate
\citet{2008arXiv0802.3848L}:
\begin{equation}
\dot{M}_{\rm
crit}^{-}=2.64\cdot10^{15}~\alpha_{0.1}^{0.01}~R_{10}^{
2.58}~M_1^{-0.85},
\end{equation}
where $\alpha$ is viscosity parameter and $R_{10}$ is disc
radius in units of $10^{10}$ cm, we obtain for the upper limit
of mass-transfer rate
\begin{equation}
\label{eq:dmdtmax}
\dot{M}_{\rm max} \raise-.5ex\hbox{$ \buildrel<\over\sim$}
2.5\cdot 10^{-7}
\left[\left(1+q\right)^{1/3}\left(0.5-0.227 \log q\right)\right]^{10.32} P_d^{1.72}f_t^{2.58}~~\rm M_{\sun}\, yr^{-1},
\end{equation}
where $P_d$ -- orbital period in days,
$f_t \raise-.5ex\hbox{$ \buildrel<\over\sim$} 0.48$ -- fractional disc truncation radius.

In Fig.~\ref{fig:pdmdt} we compare the model mass-transfer
rates with the estimates obtained by the two above mentioned
methods. The ``leaky disc'' estimates of $\dot{M}_{\rm max}$
for XTE J1118+480, GRO J0422+32, and    GRS~1009-45 strongly
suggest that the AML in short-period LMBHBs might be really
defined by GWR only.

\section{Conclusion}

We calculated models of the Galactic population of short-period
low-mass black-hole binaries which are identified with soft
X-ray transients. We found that using  the values of the
common-envelope parameter $\alpha_{ce}\lambda$ between $\approx
(0.5 - 2)$ and assuming a strongly reduced magnetic braking it
is possible to reproduce, (within the uncertainty of
observations) the number of the LMBHBs in the Galaxy and  the
effective temperatures and masses of the donors in these
systems (as inferred from the spectra of the stars). The above
mentioned values of $\alpha_{ce}\lambda$ imply that the
common-envelope expulsion in the progenitors of SXTs has to be
very efficient and that sources other than orbital energy may
be required in this process.

In our model, all short-period LMBHB systems are transient in
agreement with observations.

Model mass-transfer rates in LMBHBs are consistent with the
upper limits derived from observations under assumption that
accretion discs in SXTs are leaky.

\acknowledgements 
We thank G. Nelemans for providing models of zero-age
populations of LMBHBs and G. Dubus for the analysis of
stability of accretion disks. Both of them are acknowledged for
numerous discussions. LRY acknowledges organizers of the
conference for financial support. This study was supported by
RFBR grant 07-02-00454 and Russian Academy of Sciences Basic
Research Program ``Origin and Evolution of Stars and
Galaxies''.


\end{document}